# Effect of the regularization hyperparameter on deep learning-based segmentation in LGE-MRI

Olivier Rukundo
Department of Clinical Physiology, Lund University,
Lund, Sweden
olivier.rukundo@med.lu.se

**ABSTRACT**

The extent to which the arbitrarily selected L2 regularization hyperparameter value affects the outcome of semantic segmentation with deep learning is demonstrated. Demonstrations rely on training U-net on small LGE-MRI datasets using the arbitrarily selected L2 regularization values. The remaining hyperparameters are to be manually adjusted or tuned only when 10% of all epochs are reached before the training validation accuracy reaches 90%. Semantic segmentation with deep learning outcomes are objectively and subjectively evaluated against the manual ground truth segmentation.

## 1. INTRODUCTION

The late gadolinium enhancement (LGE) is the new principle established from the introduction of gadolinium contrast agents in cardiac magnetic resonance imaging [1]. Magnetic resonance imaging (MRI) is one of the modern medical imaging techniques which enables non-invasive assessment of cardiac structures and functions as well as diagnosis, prognosis, monitoring, and treatment of diseases [2]. After the LGE-MRI, the images produced can, with excellent reproducibility, depict the myocardium (i.e., a muscular layer of the heart, that consists of cardiac muscle cells arranged in overlapping spiral patterns) [3],[4]. The automated or deep learning-based semantic segmentation of the myocardium is one of the important steps to achieve fully automated quantification of myocardial infarction or cardiovascular diseases [5]. However, the accurate segmentation of the endocardial and epicardial boundaries of the myocardium remains very challenging due to many reasons, including low contrast, large variation in intensity and shapes, as well as methods/approaches [5]. To achieve a fast and accurate segmentation of the myocardium, the U-net is used. In brief, U-net is the convolutional neural network architecture for biomedical image segmentation [6]. The review of interest on image segmentation with deep learning is presented in [2], where authors reviewed over 100 papers on cardiac image segmentation with deep learning. Here, authors defined the goal of training as finding proper values of the network parameters to minimize the loss function and briefly discussed common loss functions as well as reducing overfitting. Also, authors pointed out overfitting as the biggest challenge of training deep neural networks for medical image analysis and several techniques developed to alleviate the overfitting problem, such as weight regularization, dropout, ensemble learning, data augmentation, and transfer learning [2]. In weight regularization, common methods used to constrain the weights include L1 and L2 regularization, which penalize the sum of the absolute weights and the sum of the squared weights, respectively [2]. In [7], L1 and L2 regularization were also discussed, with the relevant formula of the cost functions, and authors mentioned that the L2 regularization provided the maximum accuracy within their model for smaller or less complex datasets. Therefore, in this preliminary work, different L2 regularization values are arbitrarily selected to first create relevant U-nets and later evaluate the extent to which such arbitrarily selected values affect the deep learning-based segmentation outcomes. The rest of the paper is organized as follows: The second part defines the training hyperparameter and introduces the optimization strategy adopted. The third part provides information on datasets, network settings, evaluation metrics, and experimental results. The conclusion is given in the fourth part.

## 2. OPTIMIZATION OF HYPERPARAMETERS

A hyperparameter can be defined as a parameter to be set or configured before applying a learning algorithm to a given dataset [8]. Hyperparameter selection or optimization remains very challenging, especially in deep learning, because there exists no hyperparameter number that works everywhere [8]. This makes intrinsic and vague the relationship between hyperparameters and network performance [9]. There are two categories of hyperparameters relevant to the network architecture (e.g., kernel size, network width, and depth) and network training (e.g., mini-batch size, learning rate, etc.) which, if not set accurately, can worsen the performance of a deep learning network or model, in terms of accuracy and efficiency [10], [11]. Therefore, for hyperparameters optimization purposes, the author follows the simplest principle of tuning or updating hyperparameters other than L2 regularization only when 10% of all epochs are reached before training validation accuracy reaches at least 90%.

## 3. EXPERIMENTS

**A) Image datasets**: LGE-MRI datasets containing 3587 images of the size 128 x 128, were used. Such images were exported from image stacks using Segment CMR software (version 3.1.R8225), the design and validation of which were presented in [12]. Ground truth segmentation images were converted into masks with the help of the same Segment software. Each ground truth segmentation image consisted of three classes corresponding to 255, 128, and 0 labels. The dataset was split into three datasets, namely training dataset (60%), validation dataset (20%), and testing dataset (20%).

**B) Metrics and simulation software**: To evaluate the quality of cardiac images segmentation using U-net against the manual ground truth segmentation, class metrics, namely: classification accuracy, intersection over union (IoU), and mean (boundary F-1) BF score were used to (1) estimate the percentage of correctly identified pixels for each class, (2) achieve statistical accuracy measurement that penalizes false positives and (3) see how well the predicted boundary of each class aligns with the true boundary or simply use a metric that tends to correlate with human qualitative assessment, respectively [13], [14]. Additional performance measurements were also provided (See Figure 1: bottom right). The simulation software was MATLAB R2020b, but Segment software works better with MATLAB R2019b.

**C) U-net settings and graphic card**: The U-net's training hyperparameters were manually adjusted based on the observation of the training graph with the possibility for new adjustments when 10% of all epochs were reached before the training validation accuracy reached 90%. Here, U-net's training hyperparameters, manually adjusted, included the number of the epochs = 180, minimum batch size = 16, initial learning rate = 0.0001. Here, the arbitrarily chosen L2 regularization values were 0.005, 0.0005, and 0.00005, respectively. Adam was the optimizer. The execution environment was multi-GPU with Nvidia Titan RTX graphic cards. Data augmentation options used to increase the number of images in the dataset used to train the U-net were a random reflection in the left-right direction as well as the range of vertical and horizontal translations on the interval ranging from -10 to 10.

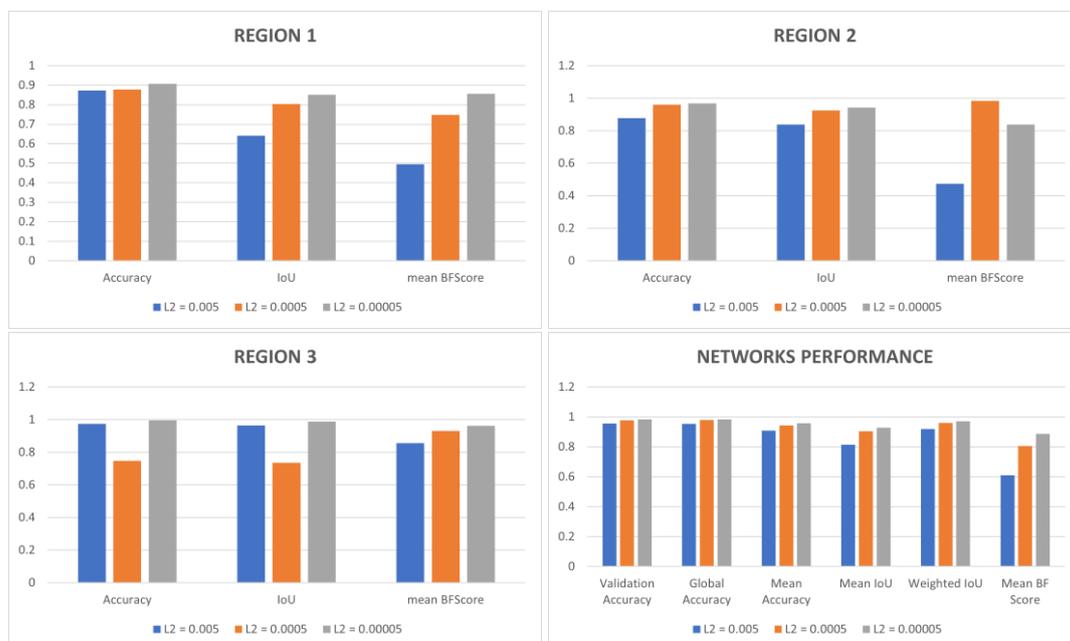

Figure 1: Performances of automated semantic segmentation of the regions adjacent to endocardial and epicardial boundaries with U-net (L2=0.005), U-net (L2=0.0005), and U-net (L2=0.00005).

**D) Results and discussions**: Figure 1 shows the results related to segmenting region 1, region 2, and region 3, which correspond to white, grey, and black colors or labels in the output masks shown in Figure 2. As can be seen, for region 1 (See Figure 1: top left), the U-net (L2 = 0.00005) or the network based on L2 = 0.00005 achieved the best accuracy, IoU, and mean BF scores. For region 2 (See Figure 1: top right), the network based on L2 = 0.00005 achieved the best accuracy and IoU scores. For region 3 (See Figure 1: bottom left), the network based on L2 = 0.00005 again achieved the best accuracy, IoU, and mean BF scores. In the bottom right side of Figure 1, the network based on L2 = 0.00005 achieved the best scores using the metrics mentioned. Referring to training graphs (not included in this work), the regularization hyperparameter L2 = 0.005, L2 = 0.0005, and L2 = 0.00005 showed that the final validation accuracy scores achieved by relevant U-nets were 95.99%, 97.23%, and 98.43%, respectively.

In Figure 2, the automated semantic segmentation with U-net is referred to as "network" while the manual ground truth segmentation is referred to as "manual". Two sample LGE-MRI images (from a test dataset), automated semantic segmentation with U-net outputs masks and manual ground truth segmentation output masks are presented in Figure 2. Here, automated semantic segmentation with U-net output masks and manual ground truth segmentation output masks as well as their differences are presented for human or subjective evaluations. Also, two test images (*sample image 1 and sample image 2*) are presented to demonstrate their relevance to output masks obtained from both segmentation approaches.

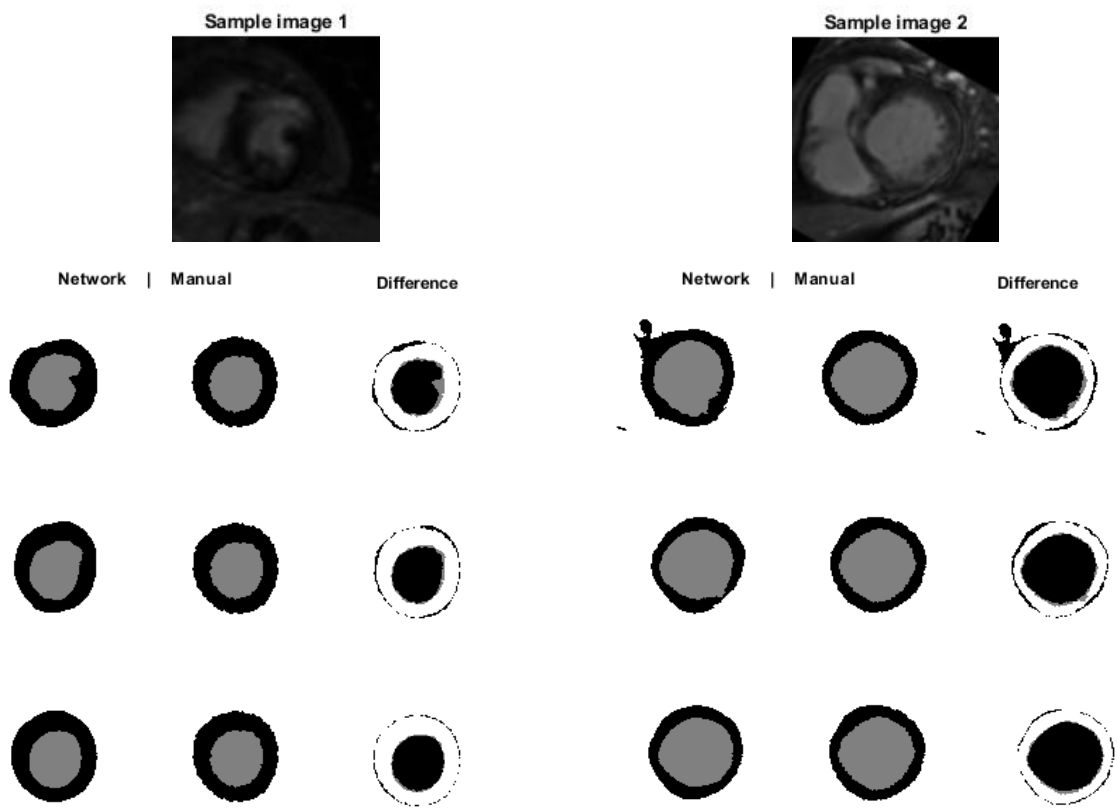

Figure 2: Differences between the automated semantic segmentation with U-nets and manual ground truth segmentation output masks.

As can be seen, under the network and manual columns, the myocardium is depicted by a black area surrounded by white and grey areas delineated by endocardial and epicardial boundaries. From top to bottom, under the network column, the first, second, and third rows present the automated semantic segmentation with U-nets or networks that used L2 = 0.005, L2 = 0.0005 and L2 = 0.00005, respectively. The difference between the automated semantic segmentation with U-nets and manual ground truth segmentation output masks is shown under the difference column. Here, with L2 = 0.005, the difference is bigger than in other cases involving the networks based on L2 = 0.0005 and L2 = 0.0005. In both two cases shown, under the two test images in Figure 2, it is visible that as the L2 values become smaller, such output masks differences become smaller too. Also, it is visible that the shape of the output mask of automated semantic segmentation with U-net using or based on the smallest L2 value is very similar to the manual ground truth segmentation output mask.

## 4. CONCLUSION

In conclusion, experiments demonstrated that the smaller L2 regularization value tended to better outcomes of semantic segmentation with U-net. In other words, the larger L2 regularization value tended to worsen the outcomes of the semantic segmentation with U-net when compared to the manual ground truth segmentation. The effect or extent of acceptable L2 regularization value could be seen based on the shapes or structures of output masks of semantic segmentation with U-nets shown in Figure 2.